\documentclass[12pt]{iopart}
\usepackage{amssymb,latexsym,epsfig}
\newcommand{\beq}{\begin{equation}}
\newcommand{\eeq}{\end{equation}}
\newcommand{\bqa}{\begin{eqnarray}}
\newcommand{\eqa}{\end{eqnarray}}
\newcommand{\fr}{\frac}

\begin{document}
\title{Unpolarized radiative cylindrical spacetimes: Trapped surfaces and quasilocal energy}
\author{S\' ergio M. C. V. Gon\c calves}
\address{Department of Physics, Yale University, New Haven, Connecticut 06511, U.S.A.}
\date{\today}
\begin{abstract}
We consider the most general vacuum cylindrical spacetimes, which are defined by two global, spacelike, commuting, non-hypersurface-orthogonal Killing vector fields. The cylindrical waves in such spacetimes contain both $+$ and $\times$ polarizations, and are thus said to be unpolarized. We show that there are no trapped cylinders in the spacetime, and present a formal derivation of Thorne's C-energy, based on a Hamiltonian reduction approach. Using the Brown-York quasilocal energy prescription, we compute the actual physical energy (per unit Killing length) of the system, which corresponds to the value of the Hamiltonian that generates unit proper-time translations orthogonal to a given fixed spatial boundary. The C-energy turns out to be a monotonic non-polynomial function of the Brown-York quasilocal energy. Finally, we show that the Brown-York energy at spatial infinity is related to an asymptotic deficit angle in exactly the same manner as the specific mass of a straight cosmic string is to the former.
\end{abstract}
\pacs{0420D, 0420J, 0430W}

\vspace{9.3cm}

Journal reference: Class. Quantum Grav. {\bf 20} 37 (2003)

\maketitle

\section{Introduction}
Cylindrically symmetric spacetimes are the simplest spacetimes with a non-trivial field content which admit exact solutions containing gravitational radiation~\cite{einstein&rosen37}. They also provide useful test beds for quantum gravity~\cite{ashtekar&pierri96-korotkin&samtleben98}, numerical relativity~\cite{inverno97}, and probes of the hoop and cosmic censorship conjectures~\cite{berger&chrusciel&moncrief95,goncalves02,goncalves&jhingan02}. The best studied vacuum case is that of the Einstein-Rosen (ER) metric~\cite{beck25}, which contains two spacelike, commuting, hypersurface-orthogonal Killing vector fields, $\xi_{(z)}\equiv\partial_{z}$ and $\xi_{(\phi)}\equiv\partial_{\phi}$, whose orbits generate translations and rotations with respect to the axis of symmetry, respectively. It has been shown that there are no trapped surfaces in the vacuum ER spacetime~\cite{berger&chrusciel&moncrief95}, and this result has recently been generalized to include the presence of matter~\cite{goncalves02}. The ER metric is a special case of the most general cylindrically symmetric vacuum metric---the Jordan-Ehlers-Kundt-Kompaneets (JEKK) metric~\cite{jekk}---whose Killing vectors are not hypersurface orthogonal, whence the cylindrical waves have two polarizations. Whether or not trapped surfaces develop in  such unpolarized metrics (without {\em a priori} asymptotic flatness conditions) has remained an open question. Here, we explicitly show that there are no trapped cylinders in the JEKK spacetime, thus in accord with the spirit of the hoop conjecture~\cite{thorne72}.

A related issue is that of the so-called C-energy in radiative cylindrical spacetimes~\cite{thorne65}. Because of the translational symmetry associated with the Killing vector $\xi_{(z)}$, it is possible to define an energy-like quantity per unit Killing length, which measures the amount of gravitational energy carried by the cylindrical waves towards infinity. Such ``energy'' is conserved for every Killing orbit of $\xi_{(z)}$, but obviously diverges in the four-dimensional spacetime. C-energy was originally defined in a somewhat heuristic manner by exploiting the local field equations, without recourse to a Hamiltonian formulation~\cite{thorne65}. Here, we derive a formal definition of C-energy (which includes the unpolarized case), by means of a Hamiltonian reduction of the field equations, wherein C-energy arises naturally from the Hamiltonian constraint. To exploit the relation between this formal definition of C-energy and the actual physical energy of the system, we compute the Brown-York quasilocal energy~\cite{brown&york93}, which is the value of the Hamiltonian that generates unit magnitude proper-time translations in a timelike direction orthogonal to spacelike hypersurfaces at some fixed spatial boundary. The C-energy is {\em not} the physical energy of the system (as per the Brown-York definition adopted herein), being instead related to it by a simple monotonic non-polynomial function. We also discuss the asymptotic behavior of the metric, corresponding quasilocal energy, and its relation to the deficit angle at infinity.

Geometrized units, in which $G=c=1$, are used throughout.

\section{Hamiltonian reduction}

We follow the Hamiltonian reduction approach of Moncrief~\cite{moncrief89}, and begin by considering a fully general $(3+1)$ metric, written as
\beq
ds^{2}=e^{-2\psi}[-\alpha^{2}dt^{2}+\gamma_{ab}(dx^{a}+\beta^{a}dt)(dx^{b}+\beta^{b}dt)]+e^{2\psi}(dz+X_{i}dx^{i})^{2},
\eeq
where $a,b=1,2$, $i=0,1,2$, and the coordinates $\{t,r,\phi,z\}$ are adopted, with $r\in{\mathbb R}_{0}^{+}$, $t,z\in{\mathbb R}$, and $\phi\in[0,2\pi]$. Since we take $\xi_{(z)}$ and $\xi_{(\phi)}$ to be Killing vectors, all the metric functions depend solely on $t$ and $r$. The triplet $\{\alpha,\beta^{a},\gamma_{ab}\}$ represents an ADM parametrization of the $(2+1)$ Lorentz metric induced on the space of orbits of the Killing vector $\xi_{(z)}$. The components $X_{i}$ control the hypersurface-orthogonality of the Killing vectors: the latter are hypersurface orthogonal if and only if $X_{i}=0$.

The JEKK metric corresponds to the coordinate and gauge choice
\bqa
&&\alpha=e^{\gamma},\; \beta^{a}=X_{t}=X_{r}=0, \; X_{\phi}=\omega(t,r)\neq0, \\
&&\gamma_{ab}dx^{a}dx^{b}=e^{2\psi}dr^{2}+r^{2}d\phi^{2}.
\eqa
We shall henceforth consider this case. The ADM action for the JEKK spacetime is (modulo integration over the two Killing coordinates, $z$ and $\phi$)
\beq
I_{\rm ADM}=\int_{\Omega} dtdr(\Pi_{\psi}\psi_{,t}+\Pi_{\omega}\omega_{,t}-\alpha{\mathcal H}-\beta^{a}J_{a}),
\eeq
where $\Omega={\mathbb R}\otimes{\mathbb R}_{0}^{+}$, $(\Pi_{\psi},\Pi_{\omega})$ are the canonical momenta conjugate to $(\psi,\omega)$, $\Pi_{\gamma}\equiv0$ and $\Pi_{r}\equiv\ln r$, and
\bqa
{\mathcal H}&=&\fr{1}{r}e^{-\gamma}\left(\fr{1}{8}\Pi_{\psi}^{2}+\fr{1}{2}e^{-4\psi}\Pi_{\omega}^{2}\right)+2(e^{-\gamma})_{,r} \nonumber \\
&&+2re^{-\gamma}(\psi_{,r})^{2}+\fr{1}{2r}e^{-\gamma}e^{4\psi}(\omega_{,r})^{2}, \\
J_{a}&=&(\Pi_{\psi}\psi_{,r}+\Pi_{\omega}\omega_{,r}+\Pi_{r}\fr{1}{r})\delta_{a}^{r}
\eqa 
are the Hamiltonian and momentum densities, respectively. The constraint equations ${\mathcal H}=J_{a}=0$ yield
\bqa
&&\gamma_{,r}=\fr{1}{4r}\left(\fr{\Pi^{2}_{\psi}}{4}+e^{4\psi}\Pi_{\omega}^{2}\right)+r(\psi_{,r})^{2}+\fr{e^{4\psi}}{4r}(\omega_{,r})^{2}, \label{hcons} \\
&&\Pi_{r}=-r(\Pi_{\psi}\psi_{,r}+\Pi_{\omega}\omega_{,r}). \label{mcons}
\eqa
The explicit dependence on the momentum variables can be removed by computing the equations of motion for the canonical variables $(\psi,\omega,\gamma)$, and using these in the constraint equations. Variation of the ADM action with respect to the relevant canonical variables gives
\bqa
\psi_{,t}&=&\fr{1}{4r}\Pi_{\psi}, \label{feq1} \\
\omega_{,t}&=&e^{-4\psi}\Pi_{\omega}, \label{feq2} \\
\gamma_{,t}&=&\fr{1}{2r}\Pi_{r}. \label{feq3}
\eqa
The constraint equations (\ref{hcons})-(\ref{mcons}) read then
\bqa
\gamma_{,r}&=&r[(\psi_{,t})^{2}+(\psi_{,r})^{2}]+\fr{1}{4r}e^{4\psi}[(\omega_{,t})^{2}+(\omega_{,r})^{2}], \label{ceh} \\
\gamma_{,t}&=&2r\psi_{,t}\psi_{,r}+\fr{e^{4\psi}}{2r}\omega_{,t}\omega_{,r}. \label{cem}
\eqa
The equations of motion for the remaining canonical variables, $\Pi_{\psi}$ and $\Pi_{\omega}$, yield [where equations (\ref{feq1})-(\ref{feq2}) were used]:
\bqa
&&\psi_{,tt}-\fr{\psi_{,r}}{r}-\psi_{,rr}=\fr{e^{4\psi}}{2r^{2}}[(\omega_{,t})^{2}-(\omega_{,r})^{2}], \label{ee1} \\
&&\omega_{,tt}+\fr{\omega_{,r}}{r}-\omega_{,rr}=4(\psi_{,r}\omega_{,r}-\psi_{,t}\omega_{,t}). \label{ee2}
\eqa
Since all coordinate and gauge conditions have been fixed, equations (\ref{ceh})-(\ref{ee2}) form a complete set. Note that the wave equations for the two degrees of freedom of the gravitational field (i.e., the functions $\psi$ and $\omega$, corresponding to the $+$ and $\times$ polarization modes, respectively) are coupled through source terms, and neither of them contains the metric function $\gamma$, which is therefore fully determined by the two polarizations at each spacetime event by quadratures, via equations (\ref{ceh})-(\ref{cem}). As we shall see below, $\gamma$ is closely related to the total gravitational energy of the system per unit Killing length $z$.

\section{Trapped surfaces}

Consider a spacelike two-surface ${\mathcal S}$ with intrinsic orthonormal basis $\{e_{(a)}^{\mu};a=1,2\}$, and let $\xi^{\mu}_{\pm}$ be tangent vector fields to future-oriented null geodesics orthogonal to ${\mathcal S}$ (where the $\pm$ denotes outgoing/ingoing families), satisfying
\beq
\xi^{\mu}_{\pm}\xi^{\pm}_{\mu}=0, \; \xi_{\mu}^{\pm}e^{\mu}_{(a)}=0, \; e^{\mu}_{(a)}e^{\nu}_{(b)}g_{\mu\nu}=\delta_{(a)(b)}.
\eeq
Following the standard Newman-Penrose formalism~\cite{newman&penrose62}, we introduce a complex null vector field $m^{\mu}$ satisfying
\bqa
&&m^{\mu}=\fr{1}{\sqrt{2}}(e^{\mu}_{(1)}-ie^{\mu}_{(2)}),\; \bar{m}^{\mu}=\fr{1}{\sqrt{2}}(e^{\mu}_{(1)}+ie^{\mu}_{(2)}), \nonumber \\
&&m^{\mu}m_{\mu}=\bar{m}^{\mu}\bar{m}_{\mu}=0.
\eqa
The vectors $\{\xi^{\mu}_{\pm},m^{\mu},\bar{m}^{\mu}\}$ form an NP tetrad, uniquely determined (up to Lorentz transformations) by
\bqa
\xi^{\mu}_{\pm}\xi^{\mp}_{\mu}&=&-m_{\mu}\bar{m}^{\mu}=1, \nonumber \\
g_{\mu\nu}&=&2\xi^{+}_{(\mu}\xi^{-}_{\nu)}-2m_{(\mu}\bar{m}_{\nu)}.
\eqa
The expansion of the null geodesic congruences defined by $\xi^{\mu}_{\pm}$ is
\beq
\Theta_{\pm}:=-\fr{1}{2} \mbox{Re} (\bar{m}^{\mu}m^{\nu}\nabla_{\mu}\xi^{\pm}_{\nu}).
\eeq
The surface ${\mathcal S}$ is said to be a trapped surface (TS) if~\cite{penrose65-68}
\beq
\Theta_{+}\Theta_{-}\geq0, \label{tsd}
\eeq
i.e., if {\em both} ingoing and outgoing null congruences are either converging or diverging at ${\mathcal S}$. The inequality saturates for the case of marginally trapped surfaces, of which the special case $\Theta_{+}=0$ corresponds to an outer marginally trapped surface (OMTS). It is well known that the existence of TS implies that of OMTS~\cite{wald84}; therefore, to show that TS do not form, it suffices to show that OMTS cannot occur. 

\subsection{Geodesic null congruences}

Introducing retarded Bondi coordinates $\{u,r,z,\phi\}$, where $u=t-r$, the JEKK metric reads
\beq
ds^{2}=-e^{2(\gamma-\psi)}(du^{2}+2dudr)+e^{2\psi}(dz+\omega d\phi)^{2}+r^{2}e^{-2\psi}d\phi^{2}.
\eeq
Because of the existing symmetries, the natural congruence of geodesics to consider consists of outgoing radial null geodesics (ORNG), whose tangent vector field is orthogonal to the $z$-axis, and carries no azimuthal dependence. We thus define the following NP tetrad for the JEKK metric:
\bqa
l_{\mu}&=&\fr{1}{2}e^{2(\gamma-\psi)}\delta_{\mu}^{u}+e^{2(\gamma-\psi)}\delta_{\mu}^{r}, \label{tet1} \\
n_{\mu}&=&\delta_{\mu}^{u}, \label{tet2} \\
m_{\mu}&=&-\fr{i}{\sqrt{2}}e^{\psi}\delta_{\mu}^{z}-\fr{1}{\sqrt{2}}e^{-\psi}(ie^{2\psi}\omega+r)\delta_{\mu}^{\phi}, \label{tet3} \\
\bar{m}_{\mu}&=&\fr{i}{\sqrt{2}}e^{\psi}\delta_{\mu}^{z}-\fr{1}{\sqrt{2}}e^{-\psi}(-ie^{2\psi}\omega+r)\delta_{\mu}^{\phi}, \label{tet4}
\eqa
where the null one-form $l_{\mu}$ is dual to the tangent vector to the ORNG, i.e., $l^{\mu}$ lies on $u=\mbox{const.}$ null hypersurfaces. Cylindrical symmetry ensures that $l^{\mu}$ has vanishing $z$ and $\phi$ components, with the remaining two being related to each other by the null condition. This still leaves scaling freedom for $l^{\mu}$, and hence any Lorentz transformation preserving the direction of $l^{\mu}$  is allowed, whereby a new (but equivalent) tetrad is obtained. A general Lorentz transformation may be divided in two classes~\cite{stewart90}:

\vspace{0.3cm}

Null rotations: $l^{\mu}\rightarrow l^{\mu}$, $m^{\mu}\rightarrow m^{\mu}+al^{\mu}$, $\bar{m}^{\mu}\rightarrow\bar{m}^{\mu}+a^{*}l^{\mu}$, $n^{\mu}\rightarrow n^{\mu}+a^{*}m^{\mu}+a\bar{m}^{\mu}+aa^{*}l^{\mu}$

\vspace{0.15cm}

Spin-boost transformations: $l^{\mu}\rightarrow\alpha^{-2}l^{\mu}$, $n^{\mu}\rightarrow\alpha^{2} n^{\mu}$, $m^{\mu}\rightarrow e^{i\theta}m^{\mu}$, $\bar{m}^{\mu}\rightarrow e^{-i\theta}\bar{m}^{\mu}$,

\vspace{0.3cm}

\noindent where $a$ is a complex-valued function, and $\alpha$ and $\theta$ are real-valued functions on the manifold. Clearly, both classes of spin basis transformations leave $l^{\mu}$ tangent to the $u=\mbox{const.}$ null surface.

The expansion scalar for the ORNG associated with $l^{\mu}$ is
\beq
\Theta_{+}=-\fr{1}{2} \mbox{Re} (\bar{m}^{\mu}m^{\nu}\nabla_{\mu}l_{\nu})=\fr{1}{4r}, \label{expsc}
\eeq
which is always strictly positive for $0\leq r<+\infty$. This result remains unchanged by Lorentz transformations, which induce the following changes in the geodesic expansion:

\vspace{0.3cm}

Null rotations: $\Theta_{+}\rightarrow\Theta_{+}+a^{*}\kappa=\Theta_{+}\;\;\;$ (since $\kappa=0$ for geodesic congruences)

\vspace{0.15cm}

Spin-boost transformations: $\Theta_{+}\rightarrow\alpha^{-2}\Theta_{+}$

\vspace{0.3cm}

\noindent Hence, $\Theta_{+}$ is always positive, and therefore {\em there are no trapped cylinders in the spacetime\footnote{We have used the fact that the absence of OMTS implies that of TS, and showed that the former cannot exist. The absence of TS can also be verified directly via the definition (\ref{tsd}), using $\Theta_{\pm}$. A straighforward computation gives $\Theta_{-}=-e^{2(\psi-\gamma)}/r$, and thus $\Theta_{+}\Theta_{-}<0$, thereby showing that there are no trapped cylinders.}.} From equation (\ref{expsc}) one sees that there is an OMTS in the limit $r\rightarrow+\infty$ (for a given value of $u$); such limiting behavior is also present in the polarized case, and can be explained due to astigmatic focusing of the ORNG at ``null infinity'' caused by the interaction with a transverse plane wave component~\cite{goncalves02}.

\subsection{Geometric definition}

For the particular case of cylindrical symmetry, one may look directly at the cylinders of symmetry induced by the space of orbits of $\xi_{(z)}$ and $\xi_{(\phi)}$, and ask under what conditions are such cylinders trapped. We are thus interested in cylinders of unit Killing length $z$, defined by 
\beq
{\mathcal C}=\{(u,r,z+\tilde{z},\phi+\tilde{\phi}): \tilde{z}\in[0,1], \tilde{\phi}\in[0,2\pi]\}. \label{cyl}
\eeq
The corresponding area spanned by the Killing vectors $\xi_{(z)}$ and $\xi_{(\phi)}$ is
\beq
A=2\pi\sqrt{|\xi_{(z)}|^{2}|\xi_{(\phi)}|^{2}-(\xi_{(z)}\cdot\xi_{(\phi)})^{2}}=2\pi r, \label{area}
\eeq
where $r$ is thus the {\em area radius per unit Killing length} $z$ (whence $[A]=[r]=L$, and not $L^{2}$). A cylinder is then said to be trapped, marginal, or untrapped if the vector $V_{\mu}:=\nabla_{\mu} r=\delta_{\mu}^{r}$ is timelike, null, or spacelike, respectively~\cite{hayward00}. In the present case:
\beq
V_{\mu}V^{\mu}=e^{2(\psi-\gamma)}\geq0.
\eeq
Therefore, provided the quantity $\psi-\gamma$ does not diverge negatively (in which case a marginally trapped surface would form), there are no trapped cylinders in the spacetime. Since $2g_{ur}=g_{uu}=-e^{2(\gamma-\psi)}$, regularity of the metric automatically ensures that $V_{\mu}V^{\mu}>0$, and hence there are no trapped cylinders. At the axis of symmetry, the requirement of local elementary flatness guarantees that the former is not trapped. The axis is regular if and only if~\cite{exactsols}
\beq
\lim_{r\rightarrow0} g^{\mu\nu}|\xi_{(\phi)}|_{,\mu}|\xi_{(\phi)}|_{,\nu}=1,
\eeq
which requires that the first derivatives of $\psi$ and $\omega$ are finite and bounded away from zero, and
\beq
\lim_{r\rightarrow0} |\psi-\gamma|<\infty, \;\;\; \lim_{r\rightarrow0} |\psi|<\infty.
\eeq
Hence, $V_{\mu}V^{\mu}>0$ along the axis, which is therefore untrapped.

\section{C-energy}

\subsection{Polarized radiation}

Consider the polarized version of the JEKK metric:
\beq
ds^{2}=-e^{2(\gamma-\psi)}(dt^{2}-dr^{2})
+e^{2\psi}dz^{2}+e^{-2\psi}r^{2}d\phi^{2}. \label{er}
\eeq
The C-energy was originally defined for this class of metrics to be $\gamma(t,r)$~\cite{thorne65}. On a given spacelike slice $\Sigma_{t}$ its value at radius $r$ equals
\beq
\gamma^{\rm pol}(t,r)=\int_{0}^{r} \partial_{\rho}\gamma(t,\rho)d\rho=\int_{0}^{r} \rho[(\psi_{,t})^{2}+(\psi_{,\rho})^{2}]d\rho, \label{ceini}
\eeq
where the $G_{tt}$ Einstein equation was used in the second equality. Clearly, $\gamma^{\rm pol}$ is always non-negative, and vanishes for flat spacetime. Remarkably, $\gamma^{\rm pol}$ satisfies two additional properties, which provide further evidence for its status as a {\em bona fide} ``energy'': (i) it can be expressed in a gauge-invariant manner in terms of the existing Killing vector fields via
\beq
\gamma^{\rm pol}(t,r)=-\fr{1}{2}\ln\left(\fr{g^{\mu\nu}A_{,\mu}A_{,\nu}}{|\xi_{(z)}|^{2}}\right),
\eeq
where $A$ is the area of a unit Killing-length cylinder, given by equation (\ref{area}) with $\omega=0$, and (ii) its coordinate gradient is related to a flux vector field,
\beq
P^{\mu}:=-\fr{1}{4}\epsilon^{\mu\nu\alpha\beta}\sqrt{-g}\gamma_{,\nu}\fr{\xi_{(z)\alpha}}{|\xi_{(z)}|^{2}}\fr{\xi_{(\phi)\beta}}{|\xi_{(\phi)}|^{2}},
\eeq
which is convariantly conserved, $\nabla_{\mu}P^{\mu}=0$. This flux vector is physical in that an observer with four-velocity $u^{\mu}$ measures a C-energy density ${\mathcal E}=P^{\mu}u_{\mu}$, and the C-energy flux across a spacelike $\Sigma$ with normal $n^{\mu}$ (satisfying $n^{\mu}u_{\mu}=0$) is ${\mathcal F}=P^{\mu}n_{\mu}$~\cite{thorne65}.

While compelling, the arguments above are merely heuristic, and the C-energy thus defined did not follow from a Hamilton-Jacobi approach, wherein energy arises in a well-defined manner as the value of the Hamiltonian that generates unit time translations orthogonal to some fixed spatial boundary~\cite{lanczos70}. We shall show here that such energy can be defined in a rigorous manner, and explicitly compute it for the JEKK spacetime. First, we note that definition (\ref{ceini}) {\em is} the Hamiltonian constraint, ${\mathcal H}=0$, for the Einstein-Rosen metric. Of course, the Hamiltonian density itself is not a physical energy density (e.g., it is not measurable), but it provides a useful guide for generalizations of C-energy to the unpolarized case, and possibly matter fields, such as Maxwell and/or Higgs. Before showing that $\gamma$ is indeed {\em related to} the total gravitational energy per unit Killing length $z$, we use the Hamiltonian constraint below to obtain the corresponding quantity for the unpolarized case.

\subsection{Unpolarized radiation}

From the Hamiltonian constraint for the JEKK metric [cf. equation (\ref{ceh})] it follows that
\beq
\gamma^{\rm unpol}(t,r)=\gamma^{\rm pol}(t,r)+\Upsilon^{2}(t,r),
\eeq
where
\beq
\Upsilon^{2}(t,r)\equiv\int_{0}^{r} d\rho \,\fr{1}{4\rho} e^{4\psi}[(\omega_{,\rho})^{2}+(\omega_{,t})^{2}].
\eeq
Two points are worth remarking: (i) the unpolarized definition reduces to the polarized one when the $\omega$ mode is turned off, and (ii) the extra piece due to the latter is manifestly non-negative, which suggests that cylindrical waves with two polarizations ``carry more energy''---a reasonable expectation. To prove that such waves are actually more efficient in taking ``energy'' away to larger radii (ultimately, to asymptotic infinity) than polarized waves, one would need to show that
\beq
\left(\fr{|\gamma^{\rm unpol}_{,t}|}{|\gamma^{\rm pol}_{,t}|}\right)_{r=r_{0}}>1,
\eeq
for any given fixed coordinate radius $r_{0}$. From equation (\ref{cem}), this is equivalent to the condition
\beq
\fr{\omega_{,t}\omega_{,r}}{\psi_{,t}\psi_{,r}}>0. \label{ineq}
\eeq
As far as the author can ascertain, the field equations alone are insufficient to determine the validity of the above inequality, for generic $\psi$ and $\omega$. There is one known exact solution (which is everywhere regular) describing unpolarized radiation, for which the above equality can be explicitly checked~\cite{piran&safier&katz86}. We note that, in the polarized case, it is possible to show that $\psi_{,t}\sim-\psi_{,r}\sim-1/\sqrt{r}<0$ for large $r$ (which implies $\gamma^{\rm pol}_{,t}<0$ and $\lim_{r\rightarrow+\infty} \gamma^{\rm pol}_{,t}=0$)~\cite{goncalvesb}. It is reasonable to expect that the same will hold for the unpolarized case, since the source term for the wave equation (\ref{ee1}) becomes negligible for large radii, whence $\psi$ behaves as in the polarized case, which in turn implies that equation (\ref{ee2}) approaches a source-free cylindrical wave equation for $\omega$. We point out, however, that these are just plausibility arguments, and by no means constitute proof that inequality (\ref{ineq}) is indeed satisfied for large $r$.

\subsection{Quasilocal energy and deficit angle}

Consider a four-dimensional spacetime $M$ which is topologically the Cartesian product of a three-manifold $\Sigma$ and a closed connected segment of the real line $I$, $M\approx\Sigma\times I$. By construction, $\Sigma$ admits only Riemannian metrics, and has spatial two-boundary $\partial\Sigma$ (not necessarily simply connected), so that the product $\partial\Sigma\times I=\,^{3}\!B \subset \partial M$ is an element of the three-boundary of $M$. The quasilocal energy associated with the spacelike slice $\Sigma$ is defined to be minus the variation of the action with respect to a unit increase in proper time separation between $\partial\Sigma$ and its neighboring surface, measured orthogonally to $\Sigma$ at $\partial\Sigma$~\cite{brown&york93}. In terms of the induced two-metric $\sigma_{ab}$ on $\partial\Sigma$, it is given by the proper surface integral
\beq
E=\fr{1}{8\pi}\int_{\partial\Sigma} d^{2}x  \sqrt{\sigma} (k-k_{0}), \label{eby}
\eeq
where $k$ is the trace of the extrinsic curvature of $\partial\Sigma$ as imbedded in $\Sigma$, and $k_{0}$ is the corresponding quantity for the {\em same} surface as embedded in a three-dimensional slice of flat spacetime (such embedding is assumed to be possible). This last term corresponds to a normalization of the ``zero of energy'' with respect to flat space. For the sake of completness, we present in the Appendix a summary of the Brown-York formalism to construct the quasilocal energy, but urge the reader to consult the original reference~\cite{brown&york93} for additional details.

Due to the $z$-translation invariance of the JEKK metric, no spacelike hypersurface defined by $t=\mbox{const.}$ can admit a closed spatial two-boundary; for this reason, we shall consider instead submanifolds which are topologically cylinders of {\em finite} length, ${\mathcal C}\approx S^{1}\times [0,1]$, whence the relevant quantities are defined {\em per unit Killing length} $z$. The two-surface of interest is then a unit Killing-length cylinder with coordinate radius $r=r_{0}$, defined on a given spacelike hypersurface $\Sigma_{t}$ by (\ref{cyl}). The Riemannian three-metric on $\Sigma_{t}$ is
\beq
h_{ij}dx^{i}dx^{j}=e^{2(\gamma-\psi)}dr^{2}+e^{2\psi}dz^{2}+2e^{2\psi}\omega dz d\phi+(r^{2}e^{-2\psi}+\omega^{2}e^{2\psi})d\phi^{2},
\eeq
which induces a two-metric $\sigma_{ab}$ on the cylinder ${\mathcal C}$:
\beq
\sigma_{ab}dy^{a}dy^{b}=e^{2\psi}dz^{2}+2e^{2\psi}\omega dz d\phi+(r_{0}^{2}e^{-2\psi}+\omega^{2}e^{2\psi})d\phi^{2}. \label{2met}
\eeq
The extrinsic curvature of ${\mathcal C}$ as imbedded in $\Sigma_{t}$ is
\beq
k_{ab}=n_{i}\left(\fr{\partial^{2} x^{i}}{\partial y^{a}\partial y^{b}}+^{(3)}\!\Gamma^{i}_{jk}\fr{\partial x^{j}}{\partial y^{a}}\fr{\partial x^{k}}{\partial y^{b}}\right),
\eeq
where 
\beq
n_{i}=\fr{^{(3)}\nabla_{i}\Phi}{|h^{ij}\,^{(3)}\nabla_{i}\Phi\,^{(3)}\nabla_{j}\Phi|}=e^{\gamma-\psi}\delta_{i}^{r}
\eeq
is the spacelike unit-normal to the cylinder ${\mathcal C}$, which is parametrically defined by $\Phi(x^{i})=r-r_{0}=0$. 

The non-vanishing components of $k_{ab}$ are
\bqa
k_{zz}&=&n_{r}\,^{(3)}\Gamma^{r}_{zz}=-e^{3\psi-\gamma}\psi_{,r}, \\
k_{z\phi}&=&n_{r}\,^{(3)}\Gamma^{r}_{z\phi}=-e^{3\psi-\gamma}(\omega\psi_{,r}+\fr{\omega_{,r}}{2}), \\
k_{\phi\phi}&=&n_{r}\,^{(3)}\Gamma^{r}_{\phi\phi}=-e^{-\gamma-\psi} r[\fr{e^{4\psi}\omega}{r}(\omega\psi_{,r}+\omega_{,r})+1-r\psi_{,r}],
\eqa
which gives
\beq
k=k_{ab}\sigma^{ab}=-\fr{e^{\psi-\gamma}}{r}. \label{kay}
\eeq

We now need to compute the corresponding quantity for ${\mathcal C}$ imbedded in a slice of Euclidean space $E^{3}$, with the standard metric
\beq
h^{\rm flat}_{ij}dx^{i}dx^{j}=dr^{2}+dz^{2}+r^{2}d\phi^{2}.
\eeq
A straightforward computation yields $k^{\rm flat}_{ab}=-r\delta_{a}^{\phi}\delta_{b}^{\phi}$, from which it follows that
\beq
k_{0}=\left(k_{ab}\sigma^{ab}\right)_{\rm flat}=-\fr{1}{r}. \label{kay0}
\eeq

From equations (\ref{eby}), (\ref{kay}), and (\ref{kay0}), the Brown-York energy is
\bqa
E&=&\fr{1}{8\pi}\int_{\mathcal C} \sqrt{\sigma}d^{2}x (k-k_{0}) \nonumber \\
&=&\fr{1}{8\pi}\int_{z}^{z+1} d\bar{z} \int_{0}^{2\pi} d\phi\, r \left(-\fr{e^{\psi-\gamma}}{r}+\fr{1}{r}\right) \nonumber \\
&=&\fr{1}{4}(1-e^{\psi-\gamma}).
\eqa
Since $E$ is evaluated for a constant time slicing $\{\Sigma_{t}: t=\mbox{const.}\}$ at some fixed coordinate radius $r=r_{0}$, from the metric form (\ref{2met}) it follows that one can always rescale the Killing coordinate $z$, such that $\psi(t,r_{0})=0$ (such rescaling is obviously dependent on the choice of slice and $r_{0}$). The final expression for the Brown-York quasilocal energy is then
\beq
E(t,r)=\fr{1}{4}\left[1-e^{-\gamma(t,r)}\right]. \label{ceby}
\eeq
This is the total gravitational energy per unit Killing length $z$ on a given spacelike slice $\Sigma_{t}$, inside a cylinder with coordinate radius $r$, for unpolarized radiation. The expression for the polarized case is the same, with $\gamma$ being given by equation (\ref{ceini}). The total energy at spatial infinity is obtained by taking the limit
\beq
E_{\infty}=\lim_{r\rightarrow+\infty} \fr{1}{4}\left[1-e^{-\gamma(t,r)}\right]=\fr{1}{4}(1-e^{-\gamma_{\infty}}), \label{einf}
\eeq
where $\lim_{r\rightarrow+\infty} \gamma(t,r)\equiv\gamma_{\infty}\geq0$. Due to the presence of the translational Killing vector $\xi_{(z)}$, the spacetime is not asymptotically flat, so $E_{\infty}$ is merely a formal analogue of the ADM mass. In fact, due to the cylindrical-wave nature of $\psi$ and $\omega$ (cf. equations (\ref{ee1})-(\ref{ee2})), under the natural assumption of absence of incoming waves from infinity, we must have~\footnote{This asymptotic behavior can be explicitly verified for the polarized case, whose general solution is given in terms of linear combinations of Bessel functions, representing oscillating modes with slowly decaying amplitude. For all such modes the wave-field $\psi$ vanishes in the $r\rightarrow+\infty$ limit; see, e.g., Ashtekar A, Bi$\check{\rm c}$\'{a}k J, and Schmidt B G 1997 Phys. Rev. D {\bf 55} 669. For the unpolarized case, a numerical study of the general solution (Piran T, Safier P N, and Stark R F 1985 Phys. Rev. D {\bf 32} 3101), as well as a specific exact solution (cf. Ref.~\cite{piran&safier&katz86} above), show qualitatively the same asymptotic behavior.}
\beq
\lim_{r\rightarrow+\infty} \psi=\lim_{r\rightarrow+\infty} \omega=0,
\eeq
from which it follows that the JEKK metric asymptotes
\beq
ds^{2}_{\infty}=-e^{2\gamma_{\infty}}dt^{2}+e^{2\gamma_{\infty}}dr^{2}+dz^{2}+r^{2}d\phi^{2}.
\eeq
Upon the rescaling 
$$
t\rightarrow \int e^{-\gamma_{\infty}}dt, \;\;\;\;\; r\rightarrow \int e^{-\gamma_{\infty}}dr,
$$
together with equation (\ref{einf}), the asymptotic metric reads
\beq
ds^{2}_{\infty}=-dt^{2}+dr^{2}+dz^{2}+(1-4E_{\infty})^{2}r^{2}d\phi^{2}, \label{asym}
\eeq
which has the form of Minkowski minus a wedge at infinity. The deficit angle at infinity is given by
\beq
\Delta\phi_{\infty}=2\pi-\lim_{r\rightarrow+\infty} \fr{\int_{0}^{2\pi} \sqrt{g_{\phi\phi}}d\phi}{\int_{0}^{r} \sqrt{g_{rr}}d\bar{r}}=8\pi E_{\infty}. \label{dangl}
\eeq
It immediately follows that the spacetime is asymptotically flat if and only if $E_{\infty}=0$, as expected. Remarkably, the metric (\ref{asym}) is {\em exactly} the metric exterior to a cosmic string with mass per unit length $\mu\equiv E_{\infty}$, when $|\mu|\ll1$~\cite{cosmic}. That is, at asymptotic spacelike infinity, a radiative cylindrical vacuum spacetime is locally indistinguishable from the spacetime of a straight cosmic string with mass per unit length $\mu\ll1$, whose pure (radiative) vacuum analogue is the Brown-York mass per unit Killing length $z$.  We remark, however, that, whereas the deficit angle for a cosmic string is related to its specific mass by $\Delta\phi=8\pi\mu$ only when $\mu\ll1$, the relation (\ref{dangl}) holds for {\em any} value of $E_{\infty}$ (which is always non-negative and bounded from above by $1/4$, in the adopted geometrized units). This is because the relation $\Delta\phi^{\rm string}_{\infty}=8\pi\mu$ is derived in the weak-field limit, which requires $\mu\ll1$~\cite{cosmic}, whereas (\ref{dangl}) is derived under the assumption of appropriate fall-off of the wave modes $\psi$ and $\omega$, wherein the upper limit for the Brown-York mass per unit Killing length remains unchanged.

\section{Concluding remarks}

The original concept of C-energy succeeded in capturing {\em some} notion of ``energy'', which is conserved for every Killing orbit of $\xi_{(z)}$, but failed to represent the {\em actual physical energy} of the system per unit Killing length, $E$, being instead monotonically related to the latter by a simple non-polynomial function. We note that the original definition, $\gamma$, was modified in an {\em ad hoc} manner by Thorne himself~\cite{comm2} to $C=(1/8)(1-e^{-2\gamma})$, to avoid the unpleasant potential blow-up of $\gamma$ on a given spacelike slice (cf. equation (\ref{ceini})). This new definition is bounded from above by $1/8$, but is still {\em not} the true energy of the system, being related to it by $C=E(1-2E)$; the two agree if and only if $C=E=0$, corresponding to the trivial case of flat spacetime.

The notion of an ``energy'' per unit Killing length $z$ arises naturally from yet another viewpoint:  in the Einstein-Rosen spacetime, the four-metric is uniquely determined by the solution of a three-dimensional Minkowski wave equation for a minimally coupled massless scalar field (our metric function $\psi$). This being the case, one can then compute the energy of the scalar field in Minkowski space (whose conservation law follows directly from the contracted Bianchi identities), and  {\em define} that as the total energy of the cylindrical waves per unit Killing length. Such dimensional reduction procedure is well known, and was used by Ashtekar and Varadarajan to derive the generator of asymptotic time translations for $(2+1)$ gravity coupled to a massless scalar field, in the context of the phase space formulation of general relativity~\cite{ashtekar&varadarajan95}. Their expression agrees precisely with that derived here for the unpolarized case (cf. equation (\ref{ceby}), which obviously includes polarized radiation as a special case). Although the motivation and methods used were quite different, it is hardly surprising that the two results agree, since they represent the same physical quantity, which ought to be unique. We point out, however, that the polarized case entails a key simplification: due to the hypersurface-orthogonality of the two Killing vector fields, there is only a single wave equation (for $\psi$) to be solved in a curved $(2+1)$ spacetime, which is decoupled from the field equations for $\gamma$. In addition, rotational symmetry implies that such wave equation is valid in the physical $(2+1)$ geometry if and only if it is so on $(2+1)$ flat spacetime. In the presence of another polarization, all these features are absent: the two Killing vector fields are not hypersurface-orthogonal, the quotient $(2+1)$ spacetime induced by the orbits of $\xi_{(z)}$ is no longer conformally flat, and neither of the two wave modes obeys a source-free cylindrical wave equation. The present method circumvented these technical difficulties by computing directly (i.e., without resorting to dimensional reduction techniques) the Brown-York quasilocal mass, which depends explicitly on geometrical quantities defined on spacelike two-surfaces homeomorphic to $S^{1}\times[0,1]$.

Finally, we remark that the Hamiltonian reduction of Sec. 2 in itself just provides a formal way to compute the C-energy, by demanding that the latter arise from the Hamiltonian constraint. This C-energy, simply the function $\gamma$ determined on a given $\Sigma_{t}$ slice by ${\mathcal H}=0$, is a mere generalization (to include unpolarized radiation) of the original definition, and, as such, does not represent the physical energy per unit Killing length $z$. The true physical energy along the Killing orbits of $\xi_{(z)}$ is given by the Brown-York quasilocal mass, which is a monotonically increasing function of the C-energy, and is bounded from above by $1/4$. The fact that the cylindrical waves carry energy in the vacuum spacetime is reflected in the geometry at infinity by the presence of a conical defect, in much the same way the line energy density of a cosmic string generates a deficit angle (anywhere in its exterior geometry, including spatial asymptotic infinity).

\ack
I am grateful to Vince Moncrief for many valuable discussions and comments. I would also like to thank Jim Isenberg, Rick Schoen, and Brian Conrey for discussions, and for the hospitality of the Stanford Mathematics Department and the AIM (American Institute of Mathematics) facility in Palo Alto, CA, where part of this work was completed. The financial support of FCT Grant SFRH-BPD-5615-2001, and NSF Grant PHY-0098084 is gratefully acknowledged.

\appendix

\section{The Brown-York quasilocal energy}

Consider a four-dimensional spacetime $(M,g_{\mu\nu})$ with three-boundary $\partial M$, endowed with a $(2+1)$ Lorentzian metric $h_{ij}$. Let $M$ be foliated by spacelike hypersurfaces $\Sigma$, with induced Riemannian three-metric $\gamma_{ij}$. It follows that $\partial\Sigma\times[t',t'']=\,^{3}\!B\subset\partial M$, where $t'$ and $t''$ denote three-boundary elements. The Einstein-Hilbert action for pure gravity appropriate for a variational principle in which the induced metric is fixed on all the elements of $\partial M$ is
\beq
\tilde{S}=\fr{1}{16\pi}\int_{M} d^{4}x\sqrt{-g}R+\fr{1}{8\pi}\int^{t''}_{t'} d^{3}x\sqrt{h}K-\fr{1}{8\pi}\int_{^{3}\!B} d^{3}x\sqrt{-\gamma}\Theta,
\eeq
where $R$ is the four-dimensional Ricci scalar, $K$ is the trace of the extrinsic curvature of $\Sigma$ as embedded in $M$, $\Theta$ is the corresponding quantity for the three-boundary $^{3}\!B$ as embedded in $M$, and $\int^{t''}_{t'}$ is a shorthand notation for $\int_{t''}-\int_{t'}$. Variation of this action yields
\bqa
\delta\tilde{S}&=&\mbox{(terms giving equations of motion)} \nonumber \\
&&+\int^{t''}_{t'} d^{3}x p^{ij}\delta h_{ij}+\int_{^{3}\!B} d^{3}x \pi^{ij}\delta\gamma_{ij}, \label{ap1}
\eqa
where $p^{ij}$ is the usual ADM gravitational momentum density conjugate to $h_{ij}$,
\beq
p^{ij}:=\fr{1}{16\pi}\sqrt{h}(Kh^{ij}-K^{ij}),
\eeq
and $\pi^{ij}$ is the gravitational momentum density conjugate to $\gamma_{ij}$ (where conjugacy is defined with respect to $^{3}\!B$):
\beq
\pi^{ij}:=-\fr{1}{16\pi}\sqrt{-\gamma}(\Theta\gamma^{ij}-\Theta^{ij}).
\eeq
To remove ambiguities arising from different possible boundary conditions on $^{3}\!B$ (all of which must be compatible with the vanishing of boundary terms in $\partial\tilde{S}$), one subtracts an arbitrary function of the fixed boundary data, $S_{0}$, such that
\beq
S:=\tilde{S}-S_{0} \label{ap2}
\eeq
is the physically meaningful action. $S_{0}$ is a functional of $\gamma_{ij}$, and thus the variation in $S$ differs from equation (\ref{ap1}) by the term
\beq
-\delta S_{0}=-\int_{^{3}\!B} d^{3}x\fr{\delta S_{0}}{\delta\gamma_{ij}}\delta\gamma_{ij}\equiv-\int_{^{3}\!B}\pi^{ij}_{0}\delta\gamma_{ij},
\eeq
where $\pi^{ij}_{0}$ is defined as the functional derivative of $S_{0}$ (therefore a function of $\gamma_{ij}$ only). If one restricts the general variation $\delta\tilde{S}$ to variation among just the classical solutions, then the action $S$ is identified with the Hamilton-Jacobi principal function $S_{\rm cl}$. Such principal function $S_{\rm cl}$ is therefore a functional of the fixed boundary data $\{\gamma_{ij}, h'_{ij}, h''_{ij}\}$, and its variation among classical solutions yields
\beq
\delta S_{\rm cl}=\int^{t''}_{t'} d^{3}xp^{ij}_{\rm cl}\delta h_{ij}+\int_{^{3}\!B} d^{3}x(\pi^{ij}_{\rm cl}-\pi^{ij}_{0})\delta\gamma_{ij}.
\eeq

Now, to construct quasilocal (i.e., defined by surface integrals) quantities, it is convenient to introduce a $(2+1)$ ADM decomposition of $^{3}\!B$:
\beq
\gamma_{ij}dx^{i}dx^{j}=-N^{2} dt^{2}+\sigma_{ab}(dx^{a}+V^{a}dt)(dx^{b}+V^{b}dt), \label{ap3}
\eeq
where $\sigma_{ab}$ is the induced metric on the spatial two-boundary $\partial\Sigma\subset\partial\, ^{3}\!B$. Motivated by the notion of energy in non-relativistic mechanics in Hamilton-Jacobi theory (simply minus the functional derivative of the classical action with respect to time), one then defines the quasilocal energy density to be:
\beq
\epsilon:=-\fr{1}{\sqrt{\sigma}}\fr{\delta S_{\rm cl}}{\delta N}=\fr{1}{8\pi}k_{\rm cl}+\fr{1}{\sqrt{\sigma}}\fr{\delta S^{0}_{\rm cl}}{\delta N}.
\eeq
The total quasilocal energy for the hypersurface $\Sigma$ is then defined as the proper surface integral of $\epsilon$ over its two-boundary $\partial\Sigma$:
\beq
E:=\int_{\partial\Sigma} d^{2}x\sqrt{\sigma}\epsilon=-\int_{\partial\Sigma} d^{2}x\fr{\delta S_{\rm cl}}{\delta N}=\fr{1}{8\pi}\int_{\partial\Sigma} d^{2}x \sqrt{\sigma}(k-k_{0}). \label{bym}
\eeq

Consider now the action (\ref{ap2}) written in canonical form, in terms of the ADM variables defined by (\ref{ap3}):
\bqa
S&=&\fr{1}{16\pi}\int_{M} d^{4}x N\sqrt{h}[R+2\nabla_{\mu}(Ku^{\mu}+a^{\mu})]-\fr{1}{8\pi}\int_{^{3}\!B} d^{3}x N\sqrt{\sigma}k-S_{0} \nonumber \\
&=&\int_{M} d^{4}x [p^{ij}\dot{h}_{ij}-N{\mathcal H}-V^{i}{\mathcal H}_{i}]-\int_{^{3}\!B} d^{3}x \sqrt{\sigma}[N\epsilon-V^{i}j_{i}], \label{ap4}
\eqa
where $u^{\mu}$ is the timelike normal to the hypersurface $^{3}\!B$, $a^{\mu}\equiv u^{\nu}\nabla_{\nu}u^{\mu}$ is the acceleration of $u^{\mu}$, ${\mathcal H}$ and ${\mathcal H}_{i}$ are the Hamiltonian and momentum densities, and $j_{i}:=(1/\sqrt{\sigma})\delta S_{\rm cl}/\delta V^{i}$ is the momentum surface density. From the action (\ref{ap4}), the Hamiltonian is explicitly determined to be
\beq
H=\int_{\Sigma} d^{3}x (N{\mathcal H}+V^{i}{\mathcal H}_{i})+\int_{\partial\Sigma} d^{2}x\sqrt{\sigma}(N\epsilon-V^{i}j_{i}). \label{ap5}
\eeq
Comparison with equation (\ref{bym}), reveals that the quasilocal energy $E$ associated with a spacelike three-slice $\Sigma$ {\em is the value of the Hamiltonian that generates unit time translations orthogonal to the boundary $\partial\Sigma$} (i.e., the value of $H$ for $N=1$ and $V^{i}=0$ on the boundary).

\end{document}